\documentclass[sigconf]{acmart}
\usepackage{algorithm}
\usepackage{algorithmic}
\usepackage{amsmath}
\usepackage{mathtools}
\usepackage{enumitem}
\usepackage{multirow}

\AtBeginDocument{%
  }

\copyrightyear{2025}
\acmYear{2025}
\setcopyright{acmlicensed}\acmConference[CHI '25]{CHI Conference on Human Factors in Computing Systems}{April 26-May 1, 2025}{Yokohama, Japan}
\acmBooktitle{CHI Conference on Human Factors in Computing Systems (CHI '25), April 26-May 1, 2025, Yokohama, Japan}
\acmDOI{10.1145/3706598.3713741}
\acmISBN{979-8-4007-1394-1/25/04}

\begin{document}

\title{\framework{}: A Drag-and-Link Video Programming Framework for Temporal Action Localization}

\author{Yuchen He}
\orcid{0009-0003-1035-4347}
\affiliation{
    \institution{State Key Lab of CAD\&CG,\\Zhejiang University}
    \city{Hangzhou}
    \state{Zhejiang}
    \country{China}
}
\email{heyuchen@zju.edu.cn}

\author{Jianbing Lv}
\orcid{0009-0000-9833-7741}
\affiliation{
    \institution{School of Software Technology,\\Zhejiang University}
    \city{Hangzhou}
    \state{Zhejiang}
    \country{China}
}
\email{lvjianbing@zju.edu.cn}

\author{Liqi Cheng}
\orcid{0009-0000-8868-5101}
\affiliation{
    \institution{State Key Lab of CAD\&CG,\\Zhejiang University}
    \city{Hangzhou}
    \state{Zhejiang}
    \country{China}
}
\email{lycheecheng@zju.edu.cn}

\author{Lingyu Meng}
\orcid{0009-0002-1662-1840}
\affiliation{
    \institution{State Key Lab of CAD\&CG,\\Zhejiang University}
    \city{Hangzhou}
    \state{Zhejiang}
    \country{China}
}
\email{kevinmeng@zju.edu.cn}

\author{Dazhen Deng}
\orcid{0000-0002-9057-8353}
\authornote{Dazhen Deng is the corresponding author.}
\affiliation{
    \institution{School of Software Technology,\\Zhejiang University}
    \city{Hangzhou}
    \state{Zhejiang}
    \country{China}
}
\email{dengdazhen@zju.edu.cn}

\author{Yingcai Wu}
\orcid{0000-0002-1119-3237}
\affiliation{
    \institution{State Key Lab of CAD\&CG,\\Zhejiang University}
    \city{Hangzhou}
    \state{Zhejiang}
    \country{China}
}
\email{ycwu@zju.edu.cn}

\renewcommand{\shortauthors}{He et al.}
\newcommand{\framework}{ProTAL}

\begin{abstract}
  
Temporal Action Localization (TAL) aims to detect the start and end timestamps of actions in a video. However, the training of TAL models requires a substantial amount of manually annotated data. Data programming is an efficient method to create training labels with a series of human-defined labeling functions. However, its application in TAL faces difficulties of defining complex actions in the context of temporal video frames. In this paper, we propose ProTAL, a drag-and-link video programming framework for TAL. ProTAL enables users to define \textbf{key events} by dragging nodes representing body parts and objects and linking them to constrain the relations (direction, distance, etc.). These definitions are used to generate action labels for large-scale unlabelled videos. A semi-supervised method is then employed to train TAL models with such labels. We demonstrate the effectiveness of ProTAL through a usage scenario and a user study, providing insights into designing video programming framework.

\end{abstract}




\maketitle

\section{Introduction}
Temporal Action Localization (TAL) is an important task within the field of computer vision, particularly for understanding and indexing long videos~\cite{DBLP:conf/eccv/EscorciaHNG16, DBLP:conf/iccv/ZhaoXWWTL17, DBLP:conf/cvpr/ChaoVSRDS18, DBLP:conf/iccv/ZengHGTRZH19, DBLP:conf/iccv/ZhaoTG21, DBLP:conf/eccv/ZhangWL22}. TAL aims to detect the start and end timestamps of specific actions and their categories~\cite{10310147}. In real-world scenarios, most videos are untrimmed, and the actions of interest may only appear in a small portion of frames. Therefore, compared to video-level action classification~\cite{DBLP:conf/cvpr/CarreiraZ17, DBLP:journals/pami/0002X00LTG19}, TAL faces the challenge of temporally localizing actions while ignoring irrelevant frames and distracting backgrounds.

With the rapid development of computer vision techniques, deep learning-based methods~\cite{liu2023endtoend, shi2023temporal, DBLP:conf/eccv/ZhangWL22} have achieved commendable results on various TAL benchmarks, such as ActivityNet-1.3~\cite{caba2015activitynet} and THUMOS14~\cite{THUMOS14}. However, training deep neural networks for TAL often requires a large amount of annotation data on specific videos, the acquisition of which incurs significant labor costs. While single-frame supervision~\cite{DBLP:conf/eccv/MaZYZKFS20, DBLP:journals/pami/YangHZLZC22} and semi-supervision~\cite{DBLP:conf/eccv/NagZSX22a} settings have been introduced to train TAL models with fewer annotations, these methods still involve a tedious annotation process, annotators are required to label each sample individually and cross-validate the results, which remains time-consuming and labor-intensive.

As a key approach in data-centric AI~\cite{zha2023datacentric}, data programming~\cite{ratner2016data, DBLP:journals/pvldb/RatnerBEFWR17} injects human knowledge into data to generate labels for model training. Although these labels can be noisy, they are crucial for the initial training of deep learning models. 
Data programming typically involves two stages: decomposition and reconstruction. During decomposition, experts use pretrained models to generate initial labels. In reconstruction, they define labeling functions to create new labels based on these initial ones. For example, in image semantic segmentation, experts might use a pretrained model to identify segments like ``transportation'' and ``water'' in a new dataset. They can then define relations between segments, such as transportation above water, to extract the label ``boat'' for model training~\cite{DBLP:journals/tvcg/HoqueHSGR23}.

Despite its success in natural language and image processing, applying data programming to TAL presents significant challenges. \textbf{First}, decomposing actions into meaningful substructures is difficult because actions in videos are spatiotemporal data, adding complexity. For example, baseball throwing involves finer actions like hip turning and hand movement, but pretrained models often lack the accuracy to localize these atomic actions.
\textbf{Second}, the spatiotemporal nature of actions makes it challenging to define labeling functions that capture detailed action dynamics. Human actions involve complex relationships between human poses and objects across frames, requiring an effective method to translate conceptual actions from users' minds into accurate labels.

To address the first challenge, we propose \framework{}, a TAL data programming framework with multiple levels of action decomposition. The framework first breaks down actions into key events, which are then defined by fine-grained visual elements extracted by computer vision modules that recognize human poses and objects frame by frame.
The design is inspired by the observation that humans can identify ongoing actions from just a few frames, thanks to discriminative cues within the action, which we refer to as key events.

To address the second challenge, we propose a drag-and-link interaction design that enables users to define key events efficiently using a graph-based visualization. Human poses and objects are mapped to nodes on a canvas, where users can drag, link, and constrain angles between key nodes to specify relations and define key events. The design also supports smooth visual transitions from real video frames to key nodes, allowing for intuitive abstraction and definition of key events.

We developed a system to implement the proposed framework and drag-and-link interaction. After uploading a video dataset, the computer vision modules will extract human poses and objects automatically. Then, users can select the videos of interest to define key events. The human poses and objects of each frame are represented as nodes and links, which are interactive and editable. Users can select specific frames as key events and complete the definition with drag-and-link interaction. The key events defined are applied to the rest of the videos, generating frame-wise action labels for the dataset.
The labels are used to train TAL models, and the models are then applied to the dataset, which accelerates the whole process of data annotation.
\framework{} also visualizes the distribution of key events across the dataset and helps further fine-tune the annotation. With several iterations, \framework{} helps users create an initial dataset for model training.
The effectiveness of our framework and interaction design was demonstrated in a practical usage scenario and a user study. The main contributions of this paper are as follows:
\begin{itemize}[leftmargin=*]
    \item We propose \framework{}, a video programming framework that decomposes complex human actions into key events and atomic elements for flexible data programming.
    \item We design an intuitive drag-and-link interaction that quickly translates user concepts into data programming rules.
    \item We implement a system of \framework{} that facilitates TAL annotation and training, demonstrating the effectiveness of our framework and interaction design.
    \item We gain insights into interactive video programming and offer lessons for designing TAL annotation systems through controlled user studies with \framework{}.
\end{itemize}

\section{Related Work}

We review previous works on TAL, interactive annotation of video data, and data programming.

\subsection{Temporal Action Localization}

Under the wave of the deep learning era, the field of TAL has undergone revolutionary development. Leveraging the robust video backbones such as C3D~\cite{DBLP:conf/iccv/TranBFTP15}, I3D~\cite{DBLP:conf/cvpr/CarreiraZ17}, and VideoMAE~\cite{DBLP:conf/nips/TongS0022}, the technology for TAL has made significant strides. Currently, TAL primarily operates under two settings: full supervision and weak supervision.

Fully-supervised TAL is the most fundamental setting, utilizing the most labeled information for model training. The earliest work can be traced back to the detection of actions by classifying sliding window proposals~\cite{DBLP:conf/cvpr/ShouWC16}.
Subsequently, the anchor mechanism was introduced to enhance the flexibility of proposal regions~\cite{DBLP:conf/bmvc/GaoYN17}.
With the introduction of TAL-Net~\cite{DBLP:conf/cvpr/ChaoVSRDS18}, the workflow of TAL was further refined, evolving the anchor mechanism into a two-stage approach.
Similarly, ActionFormer~\cite{DBLP:conf/eccv/ZhangWL22} and TriDet~\cite{DBLP:conf/cvpr/ShiZCMLT23} have enhanced TAL performance.
For weakly-supervised TAL, UntrimmedNet~\cite{DBLP:conf/cvpr/WangXLG17} is an pioneering work, consisting of a classification module and a selection module to infer the temporal boundaries of action instances.
STPN~\cite{Nguyen_2018_CVPR} introduced sparse regularization for video-level classification.
Nguyen et al.~\cite{DBLP:conf/iccv/NguyenRF19} and Liu et al.~\cite{Liu_2021_CVPR} made effective use of background segments to enhance the accuracy.
Other settings like single-frame supervision~\cite{Li_2021_CVPR, DBLP:conf/eccv/MaZYZKFS20, DBLP:journals/pami/YangHZLZC22} have been proposed to reduce annotation costs. This setting lies between fully supervised and weakly supervised, as start and end timestamps are not required for training. Instead, the model can be trained with just one annotated frame per action segment~\cite{DBLP:conf/eccv/MaZYZKFS20} or background segment~\cite{DBLP:journals/pami/YangHZLZC22}.

Regardless of the type of supervision, state-of-the-art TAL methods have achieved impressive performance across various benchmarks. However, a significant gap persists between these methods and practical applications. These models often face the problem of ``data hunger''. Training a TAL model typically requires a large-scale annotated dataset, and obtaining these annotations requires considerable costs. While weakly supervised and single-frame supervised methods can partially mitigate this challenge, the annotation process still requires manually reviewing each video, making it time-consuming and ultimately not scalable.

\subsection{Interactive Annotation of Video Data}

With the increasing demand for automatic video analysis and understanding in industries such as manufacturing, education, and sports, the high cost of video annotation has become a key barrier to applying these models. To address this challenge, researchers in the fields of human-computer interaction have proposed various interactive video annotation frameworks.
Using rules or machine learning algorithms, these frameworks significantly reduce workload, offering an effective solution.

Kurzhals et al.~\cite{DBLP:journals/tvcg/KurzhalsHSW17} utilized video segmentation algorithms to divide eye-tracking data into multiple segments and then cluster them, enabling users to annotate multiple segments simultaneously.
HistoryTracker~\cite{DBLP:conf/chi/OnoGSDS19} employed historical data and algorithms to hot-start the annotation system, allowing baseball tracking data to be generated with minimal user input.
According to the needs of racket sports analysts, EventAnchor~\cite{DBLP:conf/chi/DengWWWXZZZW21} proposed a multi-level video annotation framework that integrates computer vision algorithms and extensive domain knowledge, facilitating efficient exploration of video content. VideoModerator~\cite{DBLP:journals/tvcg/TangWWYL22} is a system developed to annotate anomalous videos, which first recommends videos through a classifier and then provides users with three different views to analyze and annotate these recommendations.
ActLocalizer~\cite{DBLP:journals/tvcg/ChenCYWKZKEL24}, tailored for TAL tasks, helps users expand single-frame annotations to full supervision by aligning action instances with a storyline-based view, thus improving the accuracy of TAL.

However, despite the significant improvements these frameworks have made in enhancing annotation efficiency, they still face challenges when applied to TAL. Firstly, although these frameworks offer well-designed user interfaces to help users understand and explore data, they are often tailored to specific tasks or scenarios. Moreover, even with these frameworks, each video still requires handling for annotation or validation, limiting scalability. It means that constructing large-scale datasets still requires substantial time and labor. Secondly, while ActLocalizer~\cite{DBLP:journals/tvcg/ChenCYWKZKEL24} presents a method that allows users to enhance supervision in datasets with single-frame annotations, it is still not suitable for scenarios where the dataset needs to be built from scratch.

\subsection{Data Programming}

Data programming offers a scalable paradigm that allows users to quickly build large datasets from scratch for model training. As one of the most promising approaches within data-centric AI, data programming injects knowledge into data in the form of user-defined labeling functions, enabling the generation of annotated data more efficiently than manually labeling each sample individually. 
Data programming was first explored in the field of natural language processing~\cite{DBLP:conf/icml/BachHRR17, DBLP:conf/aaai/RatnerHDSPR19, DBLP:conf/icml/VarmaSHRR19}.
Snorkel~\cite{DBLP:journals/pvldb/RatnerBEFWR17} enables users to provide higher-level supervision in the form of labeling functions. This approach allows for the creation of large-scale datasets without the need to meticulously manage the resulting noise and conflicts.
Ruler~\cite{DBLP:conf/emnlp/Evensen0D20} and TagRuler~\cite{DBLP:conf/www/ChoiEDH21} enable users to efficiently obtain accurate labeled data to generate labeling functions using predefined concepts and highlighting keywords, simplifying the design of labeling functions.

Researchers have been working to expand the application scenarios of data programming.
However, there are still relatively few applications in computer vision. Visual Concept Programming~\cite{DBLP:journals/tvcg/HoqueHSGR23} was the first to extend data programming to image data. This approach begins by training a self-supervised model to extract visual concepts and then offers an interactive interface that allows users to create labeling functions without writing code, enabling iterative model training. It lacks the ability to define dynamic concepts, making it unsuitable for video data. 
Additionally, VideoPro~\cite{10292616} applies data programming to video data through sequence pattern mining, but fails to provide temporal annotations for actions, limiting its utility in TAL. 
To address these limitations, we propose a novel framework that extends data programming to TAL, aiming to bridge the gap between TAL methods and practical applications.

\section{Problem Formulation}

We first introduce the concepts of data programming and how we formulate the problem of data programming in the TAL scenario.

\textbf{Data Programming Paradigm.} To begin, we introduce the paradigm of data programming, which usually consists of two stages.
The first stage involves the automatic extraction of visual elements. Advanced computer vision algorithms are used to extract visual elements that may serve as candidates for the definition of new labels. 
The second stage focuses on defining the rules that can be used to compose the candidates together and generate new labels.

The key to effective data programming in TAL is to extract basic action elements and reconstruct them. In this study, we first decompose actions into key events inspired by the concept of ``key frames'' in video editing, which define the start and end points of transitions or animations. While key points can anchor human actions, we use the term ``key events'' instead of ``key frames'' because a key event can span several frames. This flexibility accounts for slight variations in the same action across different videos, where a single key frame would be too restrictive.
Key events are considered the bridge between the target actions and basic visual elements.

\textbf{Key Event.} A key event is an atomic event within an action characterized by changes in the relations between several visual elements, which is easier to decompose and define. For example, the ``clean and jerk'' action includes a key event $K_{0}$ (\autoref{fig:framework}E): ``The barbell moves from below the athlete's head (\autoref{fig:framework}E1) to above the athlete's head (\autoref{fig:framework}E2).''

Key events serve as anchor points for the actions, but another unresolved problem is how to define and refine the key events using low-level visual elements.
Taking the case in \autoref{fig:framework} as an example, $K_{0}$ involves two visual elements: the ``barbell'' and the ``person's head,'' with $K_{0}$ being defined by the relative position change between these two visual elements.
However, to leverage these visual elements to define key events, two key questions remain to be addressed:
\begin{itemize}
    \item [\textbf{Q1}] What \textbf{visual elements} should be extracted for the definition of key events?
    \item [\textbf{Q2}] What \textbf{constraints} are required to define a key event with visual elements?
\end{itemize}

\begin{figure*}[t]
    \centering
    \includegraphics[width=0.8\linewidth]{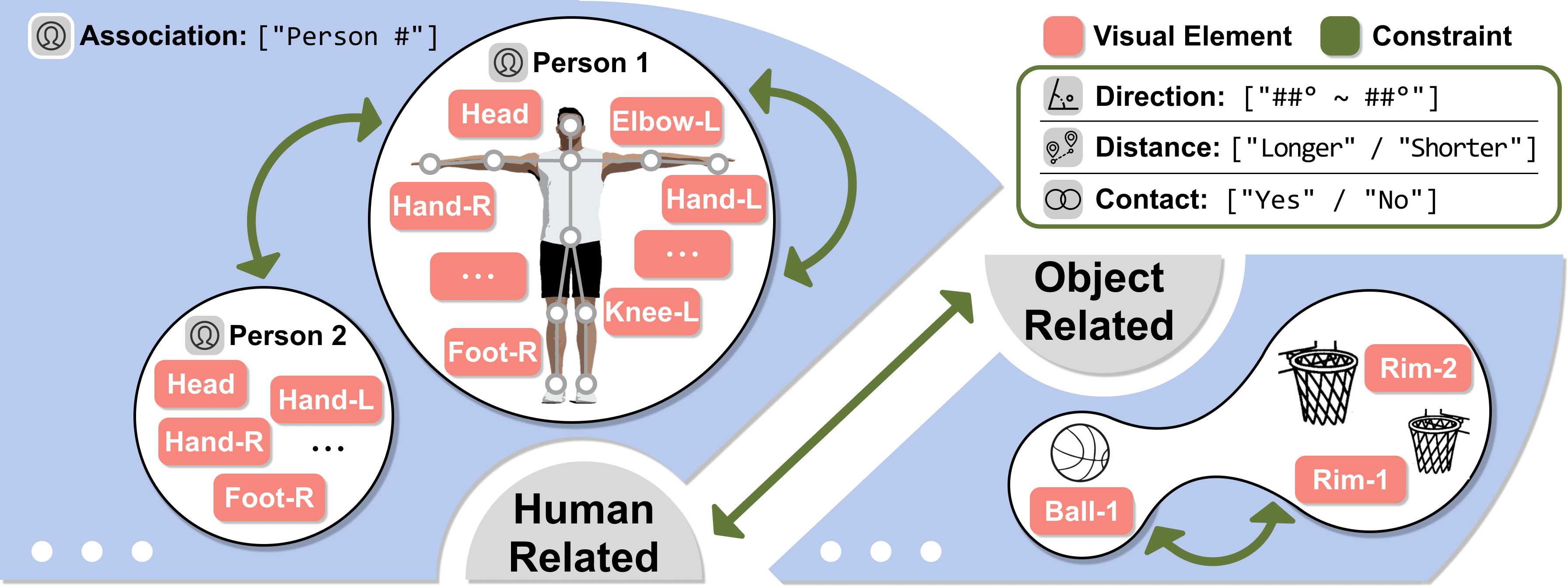}
    \caption{The space of visual elements and constraints in key event definitions. Visual elements include two categories: human-related visual elements, mainly human body parts, usually represented as skeletons; and object-related visual elements, including objects involved in the action. Constraints include direction, relative distance, contact, and association constraint.}
    \Description{This figure illustrates the space of visual elements and constraints. There are two categories of visual elements: human-related visual elements and object-related visual elements, where the human-related visual elements involve various body parts. For constraints, direction, relative distance, contact, and association constraint should also be provided.}
    \label{fig:space}
\end{figure*}

\section{Design Considerations of \framework{}}
To prepare for the design of \framework{}, we conducted a literature review and a workshop study\footnote{The study has been approved by State Key Lab of CAD\&CG, Zhejiang University.} to identify the space for visual elements and constraints.

\subsection{Literature Review}

Since a key event is a temporal and spatial substructure of an action, understanding the visual elements involved in key events requires first identifying the visual elements associated with actions. To explore this, we conducted a literature review to gain insights from previous research on human action recognition and detection. We reviewed 23 studies~\cite{DBLP:journals/ivc/AntounA23, DBLP:journals/jvcir/KhaireK22, DBLP:journals/cviu/StergiouP19, yan2018spatial, DBLP:journals/mta/BeddiarNSH20, DBLP:journals/pr/DangMWPLM20, DBLP:conf/cvpr/Duan00LD22, faure2023hoslistic, Paoletti2020skeleton, rajasegaran2023, yao2010mutual, DBLP:conf/cvpr/WangYDK0S20, DBLP:conf/cvpr/GkioxariGDH18, DBLP:conf/cvpr/XuWLZK19, feng2021relation, gu2018ava, kim2021endtoend, LIU2023109560, DBLP:conf/cvpr/PanCSLS021, zeng2022pami, zhao2020video, DBLP:journals/ijcv/ZhaoXWWTL20, Zou_2021_CVPR} and identified two main categories of action-related visual elements.
Based on the interactions involved in the actions, actions can be categorized into three categories: (1) \textbf{single-human actions}, (2) \textbf{human-human interaction}, and (3) \textbf{human-object interaction}, and focusing on these categories, many studies have tried to improve the action detection or recognition performance.
It is worth noting that object-object interactions could be considered a separate category, but they are beyond the scope of this discussion. In practice, objects can be highly complex. For instance, a modern car can be broken down into components like pistons, crankshafts, and valves, each operating with a distinct mechanism of motion. Providing a formal definition that encompasses all types of objects is inherently challenging.
Additionally, a human pose can be viewed as a simplified representation of a machine. If the target object is clearly defined, the proposed method for modeling human-human interactions could be adapted and extended to handle such scenarios.
Therefore, we focus on the discussion on single-human actions, human-human interaction, and human-object interaction.
In these categories, the interaction subjects considered are actually humans and objects. Therefore, we can start from these two interaction subjects and consider the visual elements related to the action: human-related visual elements and object-related visual elements.

\textbf{Human-related Visual Elements}. Human-related visual elements are central to actions, as the human body plays a leading role in action involving multiple body parts. Among the 23 studies, 19 utilize pre-recognized human bodies as input, with 10 in the form of poses and 9 in the form of bounding boxes. Therefore, when considering human-related visual elements, it is essential to account for the various parts of the human body.

\textbf{Object-related Visual Elements}. In the context of human-object interaction, out of 21 studies that addressed this area, 11 utilized bounding boxes of relevant objects as input, in addition to learning representations directly from RGB. Thus, for object-related visual elements, we need to focus on objects that are relevant to the action being performed.

\subsection{Workshop Study}

Through our literature review, we identified the potential types of visual elements involved in key events and answered Q1. The next step is to determine the types of relations between them should serve as constraints in key event definitions. As the concept of a key event is newly introduced in this paper, it may not be appropriate to apply element relations considered in existing action-related works.

\begin{figure*}[t]
  \centering
  \includegraphics[width=\linewidth]{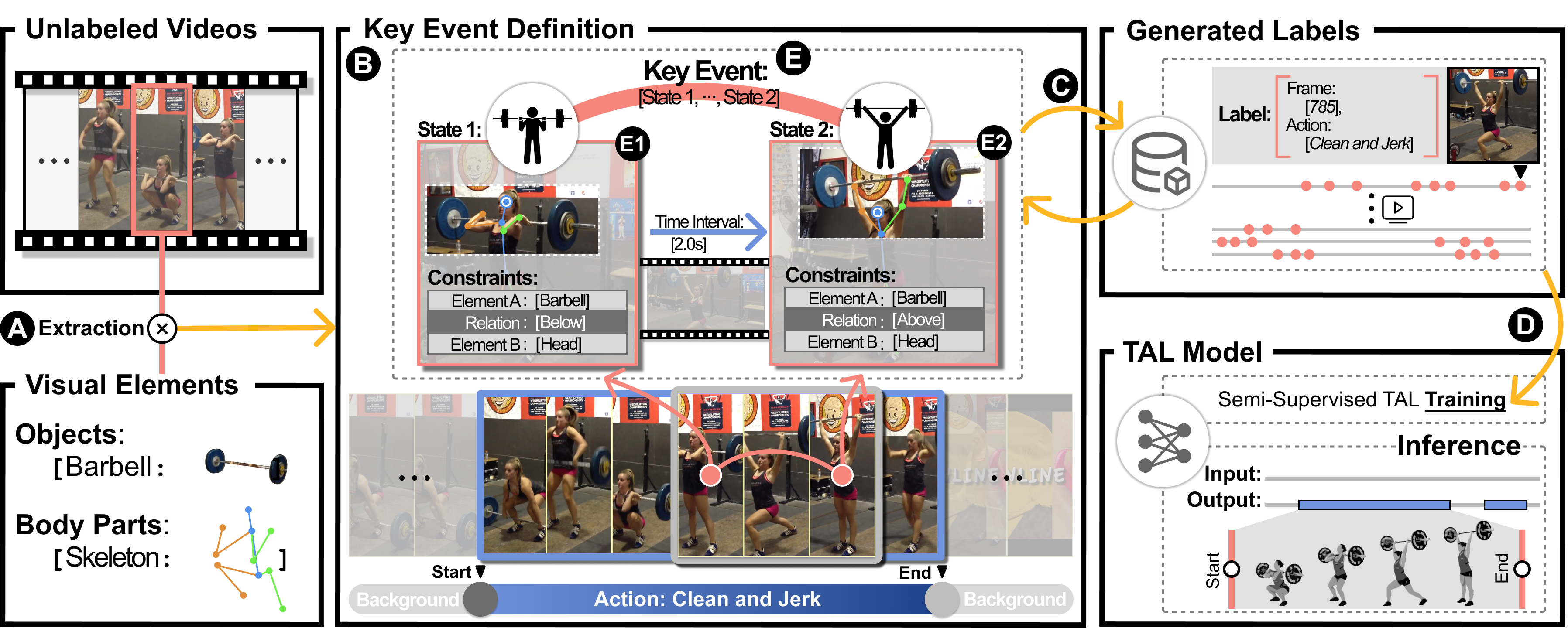}
  \caption{Framework of \framework{}. The first stage is (A) the automatic extraction of action-relevant visual elements. The second stage is (B) the defining of key events based on interactions, followed by (C) the generation of key event labels. The third stage is (D) the model training with a semi-supervised TAL method based on the generated labels.}
  \Description{This figure illustrates the proposed framework. It consists of three stages, including the automatic extraction of action-relevant visual elements, the defining of key events based on interactions followed by the generation of key event labels, and the model training with a semi-supervised TAL method based on the generated labels. A key event is a substructure within an action characterized by changes in the relations between several visual elements. For example, the ``clean and jerk'' action includes a key event: ``The barbell moves from below the athlete's head to above the athlete's head.''}
  \label{fig:framework}
\end{figure*}

A key event is characterized by changes in the relations between several visual elements, which can be represented as a series of \textbf{state} transitions. As illustrated in \autoref{fig:framework}, states 1 and state 2 correspond to two distinct states within the key event $K_0$, allowing $K_0$ to be expressed as $K_0 \coloneqq state_1 \rightarrow state_2$.
It is apparent that each state, such as states 1 and 2, can be represented by a frame in the action, indicating that a key event is, in fact, a dynamic concept composed of a sequence of static states. 
Therefore, when defining a key event, we are essentially defining a series of static states. Therefore, the relations between the visual elements in these states are also static. This strategic decomposition of key events significantly simplifies their retrieval, as it only requires identifying static frames that match the specified rules.

The nature of key events guides us in further exploring the constraint space. Following this, we conducted a workshop study with a brainstorming session and a follow-up seminar to derive the space of the constraint in detail.

\textbf{Participants.} We conducted the workshop with 8 action annotators (E1-E8) who have participated in action annotation more than 5 times and have backgrounds in programming and AI. Among them, E2 and E7 (both male) are Ph.D. in computer science, while the others are graduate students (4 in computer science and 2 in sports science, male=4, female=2). All participants have experience in action annotation for racket sports (e.g., tennis, table tennis, badminton), 75\% have experience with other ball sports (e.g., basketball, football, volleyball), and 50\% have annotation experience with other types of actions.

\textbf{Procedure.} We began by assessing the participants' backgrounds and understanding the types of action they had previously annotated. Next, we introduced the concept of key events and the visual element space derived from our earlier research. After ensuring that the participants had a good understanding of the relevant concepts, we organized a brainstorming session in which each participant was shown three videos: one containing jumping jacks (single human action), one containing handshake (human-human interaction), and another containing clean and jerk (human-object interaction). Each video contains more than 10 action instances. These actions involve multiple types of relations, including those between human-related elements, object-related elements, and between human-related and object-related elements, effectively covering all possible pairings of element types. These actions are also common to minimize potential bias due to varying levels of familiarity and to facilitate broader discussions. Participants were asked to propose a key event for each action and, assuming they had access to the bounding boxes of all action-related visual elements in the frames, provide a pseudocode (or a natural language description) that could be used to retrieve the frames corresponding to the key event, each video for 20 minutes. Following this, we held a seminar where participants summarized the 24 pieces of pseudocode and identified the type of constraints needed to define key events.

\textbf{Findings and Discussions.} All participants highlighted the importance of relative position between visual elements. Given that each frame naturally provides the bounding box position of visual elements, relative position becomes a key consideration when defining relations between them. Also, relative position is a very intuitive relation for a pair of visual elements. To express the relative position, such as ``above,'' ``to the left,'' ``upper right,'' ``upper left,'' etc., 83\% of the pseudocode examples calculated the direction angle, while 58\% involved directly comparing $x$-coordinates or $y$-coordinates. During the seminar, it was agreed that while direct coordinate comparison might be feasible for simpler direction relations such as ``above'' and ``below'', calculating the direction angle offers broader coverage and greater accuracy.

In addition to direction, participants also mentioned distance as a crucial aspect of relative position. E1 and E4 noted that using absolute pixel distance is impractical, as variations in camera shooting distance and changes in viewing angle can cause this value to fluctuate, so they opted for relative distance, comparing the magnitude of the distances between pairs of visual elements. Participants noted that direction and relative distance together were sufficient to describe a relative position. Furthermore, these relations can be applied between any type of visual element.

Beyond relative position, it was observed that in the pseudocode for the second action, all participants utilized the intersection of the bounding boxes of two individuals' hands. Participants agreed that contact is a required constraint, and the overlapping of regions can capture this relation better than distance because objects vary in size and shape. Furthermore, E2 proposed that the association constraint, which defines the relationship between body parts and their respective individuals, is essential. This association can be derived from the extracted human poses. All participants agreed that in scenarios involving multiple individuals, accurately associating body parts with the correct individuals is critical for identifying and defining key events.

\subsection{Design Principle}

Our goal is to design a TAL data programming framework that allows users to define key events through interaction and use these rule-based definitions to generate labels for unlabeled video sets to train the TAL model. Based on the previous research, we now have a clear understanding of the space of visual elements and constraints, shown as \autoref{fig:space}.

\textbf{Visual elements}. There are two categories of visual elements to consider: human-related visual elements and object-related visual elements, where the human-related visual elements involve various body parts. Therefore, when implementing the system, it is essential to provide:
\begin{itemize}
    \item [\textbf{P1}] Automatic extraction of visual elements in frames, including human body parts and action-related objects.
    \item [\textbf{P2}] Supporting direct manipulation \cite{hutchins1985direct} of the visual elements on the user interface.
    \item [\textbf{P3}] Providing intuitive visual mapping of visual elements from video frames to canvas.
\end{itemize}

\textbf{Constraints}. For constraints, it is necessary to provide the relative position relations, including direction (angle) and relative distance. In addition, contact relations, which indicate whether two visual elements are in contact, and the association constraint, which constrain the person to whom a human-related visual element belongs, should also be provided. Therefore, the design principles for constraints include:
\begin{itemize}
    \item [\textbf{P4}] Providing sufficient constraint candidates, including direction, relative distance, contact, and association constraint.
    \item [\textbf{P5}] Supporting interactive setting of constraints to define key events, with visualization of constraints on the user interface.
    \item [\textbf{P6}] Enabling users to get feedback on the generated labels and iteratively fine-tune the constraints they set.
\end{itemize}

\section{Framework of \framework{}}\label{sec:framework}

We propose \framework{}, a data programming framework designed for TAL. Built on the data programming paradigm, \framework{} incorporates the unique characteristics of temporal action data in TAL. The framework allows users to efficiently generate training labels for unlabeled videos through interaction. As shown in \autoref{fig:framework}, \framework{} follows a three-stage pipeline, which is described in detail below.

\subsection{Extraction of Action-Related Visual Element}\label{sec:fiststage}

For unlabeled video data, the first stage of \framework{} involves extracting action-related visual elements from each frame. These visual elements are then used to filter frames that contain a specific set of elements that meet defined constraints. According to P1, visual elements extracted are categorized into two groups: human-related elements and object-related elements. Using advanced computer vision models, both categories can be extracted automatically and efficiently.

For human-related elements, it is necessary to extract various body parts of the human and to distinguish which person these elements belong to (P4). Existing human pose estimation methods, such as ViTPose~\cite{DBLP:conf/nips/XuZZT22} and RTMPose~\cite{DBLP:journals/corr/abs-2303-07399}, can be employed to obtain skeleton information from each frame, thereby capturing the location of different body parts for each person in the frame. For object-related elements, state-of-the-art object detection and semantic segmentation models are highly effective in detecting or segmenting specific objects in videos, thus providing the necessary location information (P4). In addition to these models, recent advances in multimodal models, such as Grounding DINO~\cite{liu2023grounding} and Grounded SAM~\cite{ren2024grounded}, combine the strengths of various types of models to enable more robust detection and segmentation of complex visual elements using natural language prompts. These models are also useful for extracting object-related elements from videos. \framework{}'s design allows flexibly integrating computer vision models that best suit users' needs. For instance, users can integrate a detection model that is purposely trained for tennis balls to extract their positions more precisely compared to using general vision large models, such as Grounding DINO.

At this stage, \framework{} has extracted action-related visual elements from the initial unlabeled video set. For any given frame $f$ in any video $v$ within the video set, the visual elements extracted from $f$ are denoted as $ELM_{f}$:
\begin{align}
    ELM_f &= \{e_1, e_2, \cdots\, e_n\}, \\
    e_i &\coloneqq \{Type, Position, Association (\text{human-related}) \nonumber\\
    &\quad\quad ,\cdots\}, i \in \{1,\cdots,n\}, \label{equ:elm_extracted}
\end{align}
where each $e$ represents a visual element, which includes attributes such as location and category.

\subsection{Key Event Definition and Label Generation}

After the automatic extraction of visual elements, the second stage involves defining key events through an interactive interface. These key event definitions serve as rules for identifying frames that correspond to the key events and assigning labels to them. The labels are then presented to the users, enabling them to refine the key event definitions to improve label quality.

\subsubsection{The Concept of Key Event.}

In order to define a key event, denoted by $K$, users are required to specify $n_{s}$, the number of states that comprise $K$:
\begin{equation}
    K \coloneqq state_1 \rightarrow state_2 \rightarrow \cdots \rightarrow state_{n_{s}},\label{equ:eventtostate}
\end{equation}
and the threshold $thr$ of time interval between adjacent states:
\begin{equation}
    state_k \xrightarrow{t \leq thr_{k, k+1}} state_{k+1}.\label{equ:interval}
\end{equation}
For each state, users are required to provide a detailed definition. In order to define $state_k$, users are required to specify the visual elements involved, the attributes of each of these elements, and the relations between them:
\begin{align}
    state_k &\coloneqq \{ELM, REL\}, \\
    ELM &= \{e'_1, e'_2, \cdots, e'_{n_e} \},\label{equ:elementinevent}\\
    e'_i &\coloneqq \{Type, Association (\text{human-related})\nonumber\\
    &\quad\quad , \cdots \}, i \in \{1, \cdots, n_e\},\label{equ:elementattr}\\
    REL &= \{r'_{i,j}, \cdots \}, i, j \in \{1, \cdots, n_e\},\label{equ:relationinevent}\\
    r'_{i,j} &\coloneqq \{Value_1, Value_2, \cdots \}, i, j \in \{1, \cdots, n_e\},\label{equ:relationattr}
\end{align}
where $e'_i$ denotes the element $i$ involved in the state definition, $n_e$ denotes the number of such elements, and $r'_{i, j}$ denotes the user defined relation between element $i$ and element $j$. The set of values $\{Value_1, Value_2, \dots\}$ corresponds to the specific parameters or attributes for the corresponding type of relation.

\subsubsection{The Retrieval of Key Event Frames}

When users complete the definition of a key event, the frames in the videos that match the user-defined key event definition will be retrieved and assigned labels. Specifically, in each state within the key event, the visual elements and the constraints together serve as the rules for searching through each frame in the video to identify those that align with the state's definition. After retrieving the frames corresponding to each state, the sequence of frames that meet the conditions based on the user-defined time interval threshold $thr$ between the states represents the frames of the key event. Thus, retrieving key event frames in the video primarily involves retrieving frames that satisfy the definitions of each state within the key event. First, we represent frames in the video abstractly. Given all visual elements $ELM_{f}$ extracted from a frame $f$ and all computable relations between them $REL_{f} = \{r_{i, j}, \cdots \}$, $f$ can be structurally represented as a graph, denoted as $G_f \coloneqq \{ELM_f, REL_f\}$, since each visual element can be treated as a node with attributes and each relation between a pair of nodes can be considered as an edge with weights. This structure aligns with the state definition $G_{state_k} \coloneqq \{ELM, REL\}$.

Given the state definition $state_k$, as $ELM_f$ may contain redundant visual elements, determining whether $f$ is a frame corresponding to $state_k$ requires checking if $G_{state_k}$ is a subgraph of $G_f$. This means that determining whether a frame satisfies the state definition is essentially a subgraph matching problem with edge weights.

Since state definitions are generally not overly complex and the number of nodes in the subgraph is typically small, a search algorithm with pruning, denoted as $\Phi$, can be employed for subgraph querying:
\begin{align}
    \Phi (G_f, G_{state_k}) = \left\{ \begin{aligned}
        & True \quad &\text{$f$ corresponds to $state_{k}$},\\
        & False \quad &\text{otherwise.}
    \end{aligned} \right.
\end{align}
When $\Phi (G_f, G_{state_k}) = \text{True}$, the frame $f$ corresponds to $state_{k}$; otherwise, it does not. After labeling all frames corresponding to each key event, the results are presented to the users, guiding them to refine the key event definitions in order to generate more accurate labels for TAL training.

\subsection{TAL Model Training}

After completing the first two stages, the original video dataset now contains sparse frame-wise action labels. The objective of this stage is to utilize these frame labels to train the TAL model.

\subsubsection{Problem Statement}

Given a video $v$ with $T$ frames, with an action instance in $v$ from $[t_l : t_r]$, where $0 \leq t_l \leq t_r \leq T$. Since the key event is a substructure of the action, the frames labeled by states of a key event lie within the action. The generated labels for the action instance consist of several frames between $t_l$ and $t_r$, denoted as $Label_{\framework{}} = \{t_1, t_2, \cdots, t_m\} \subseteq \{t_l, \cdots, t_r\}$, with each labeled frame implicitly assigned an additional state label. This differs from full supervision labels, $Label_{full} = \{t_l, \cdots, t_r\}$, which include all frames within the action instance, and from the single-frame supervision labels used in SF-Net, $Label_{SF} = \{t'\}, t'\in \{t_l, \cdots, t_r\} $, where only one frame within the action instance is labeled. Furthermore, in both full supervision and single-frame supervision, every action instance is assigned labels. For \framework{}, however, there may be instances that remain unlabeled.

\subsubsection{Training Method}

\framework{} employs a semi-supervised approach by extending SF-Net to train with $Label_{ProTAL}$. SF-Net can be trained with any number of frame labels, but cannot fully leverage unlabeled samples for representation learning. To address this, we refine the classification target of the classification head to the state level. Given that the states within key events are inherently ordered, a state order loss on unlabeled videos is introduced during training to penalize any incorrect prediction of state order.

\begin{figure*}[t]
    \includegraphics[width=\linewidth]{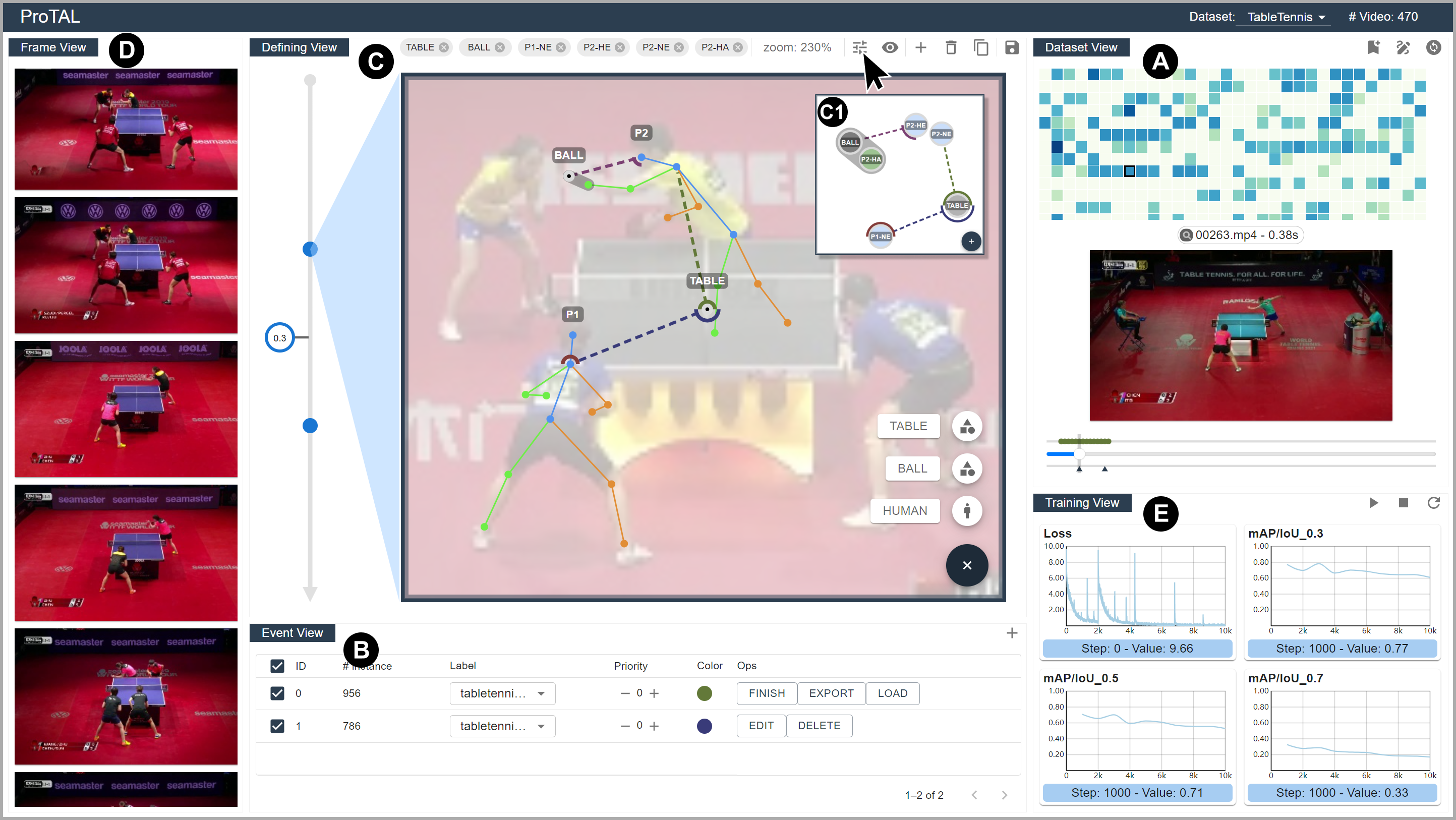}
    \caption{System screenshot. Users can navigate the video dataset and identify key events in \textit{Dataset View} (A). They can add key events in \textit{Event View} (B) and define them through drag-and-link interactions in \textit{Defining View} (C). The distribution of generated labels and the labeled frames can be reviewed in \textit{Dataset View} and \textit{Frame View} (D) to guide the refinement of definitions. \textit{Training View} (E) shows the progress of TAL model training based on the generated labels.}
    \Description{This figure is a screenshot of our system. The user interface consists of five views: Dataset View, Event View, Defining View, Frame View, and Training View. Dataset View provides an overview of the dataset with a cell matrix and a video display module. Event View facilitates the management of key events, allowing users to add, delete, export, or import events. Defining View features a canvas for defining key events. Frame View lists frames retrieved based on user-defined rules. Training View shows the status of model training. The system also includes a mode switch function that displays the key event definition in an alternate view.}
    \label{fig:screenshot}
\end{figure*}
  
\section{Interface Walkthrough: A Practical Scenario}\label{sec:ui}

Based on the proposed \framework{} framework in \autoref{sec:framework}, a prototype system with a drag-and-link interactive user interface was implemented, as shown in \autoref{fig:screenshot}. In this section, we present a practical usage scenario where the system is used to program an unlabeled table tennis video dataset for TAL training. We demonstrate how the user interact with the system throughout the process and evaluate the final TAL model.

\textbf{Background.} \textit{Alex} is a data analyst with extensive experience in annotating table tennis action data. He has participated in the annotation of a number of table tennis-related datasets and is proficient in the use of AI methods to identify objects such as balls, players, tables, and actions in video. The current method for segmenting rallies in table tennis match videos relies on identifying score changes on the scoreboard. However, this approach is sometimes inaccurate due to the delays in score adjustment during broadcast. To address this issue, Alex aims to train a TAL model that can temporally locate table tennis serve actions, with the objective of refining rally segmentation by detecting the time intervals of serve actions.

\textbf{Implementation Details.} The prototype system uses several computer vision modules to extract visual elements. For human-related elements, RTMPose~\cite{DBLP:journals/corr/abs-2303-07399} is integrated to extract human poses from videos. For object-related elements, such as the ball and the table, an off-the-shelf detection model trained specifically for table tennis analysis tasks is utilized. According to \autoref{equ:elm_extracted}, the extracted attributes of visual elements include position, type, and association (derived from human poses).
To track individuals and objects across different states of a key event instance, we use the Intersection over Union (IoU) of bounding boxes of adjacent frames, given the short time span. This ensures that when matching subgraphs, individuals with the same ID in each $G_{state}$ correspond to the same person in the video.
For relative distance, during subgraph matching, we ensure that the length order of each corresponding edge pair remains consistent with the definition.
For the contact constraint, two bounding boxes are considered to be in contact if their IoU exceeds a predefined threshold.

\subsection{User Interface Overview}

\textbf{Functionality.} The user interface includes five views. \textit{Dataset View} supports video browsing and label review. \textit{Event View} allows key event management. \textit{Defining View} displays a canvas for defining key events. \textit{Frame View} lists the frames retrieved based on the user-defined key events. \textit{Training View} displays the status of model training.

\textbf{Interaction.} The drag-and-link interaction design is inspired by motion editing techniques in animation. In animation editing, keyframes are often manipulated by dragging human joint points to create or adjust motion sequences, as demonstrated in systems like TimeTunnel~\cite{timetunnel} and the pin-and-drag interface~\cite{pinanddrag}. Additionally, \framework{} abstracts each state within a key event as a graph, making drag-and-link interactions a natural fit for defining states. Dragging provides an intuitive way to adjust nodes~\cite{WANG202149} or subgraphs~\cite{renoust2019animated} within the graph, while link is an inherent component of the graph~\cite{HAN202161, Murakami2025, Li2025graph}, effectively representing the relations between nodes. This design ensures that defining relations between nodes through linking is intuitive.

\subsection{Data Programming on Table Tennis Videos}

\begin{figure}[ht]
  \centering
  \includegraphics[width=\linewidth]{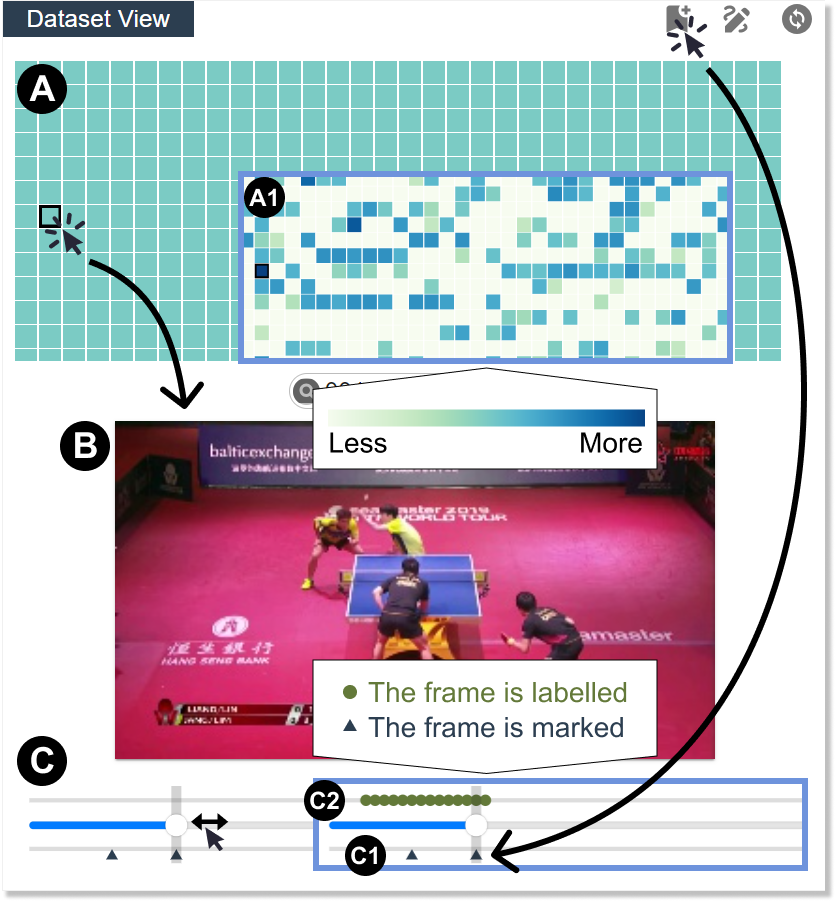}
  \caption{The \textit{Dataset View} contains: (A) a cell matrix, where each cell represents a video, (B) a video display module, and (C) a timeline module containing two timelines, the top one (C2) showing the label distribution and the bottom one (C1) showing the user's markers.}
  \Description{This figure illustrates the Dataset View in the user interface, with a cell matrix, where each cell represents a video and the color encodes the number of labels in the video, a video display module, and a timelines module. The timeline module contains two timelines: the top timeline displays the label distribution, while the bottom one shows the user's markers.}
  \label{fig:datasetview}
\end{figure}

\subsubsection{Dataset Browsing and Frame Marking}

Alex started with a dataset of 470 unlabeled table tennis video clips. The system first completed the extraction of the visual element information.

\textbf{Video Browsing}. Alex began by using the \textit{Dataset View} (\autoref{fig:datasetview}) to get an overview of the videos. The \textit{Dataset View} presents a cell matrix (\autoref{fig:datasetview}A), where each cell represents a video. By clicking on a cell, the video display module (\autoref{fig:datasetview}B) below displays the corresponding video. The timeline module (\autoref{fig:datasetview}C) includes a draggable progress bar to control the playback of the video and two parallel auxiliary timelines. Alex clicked on several videos to get a general sense of the dataset.

Drawing from his experience in table tennis data annotation, Alex believes that the serving action is distinct from other strokes because it ``\textit{involves a ball-throwing event}.'' Therefore, he considered using this ball-throwing event as the blueprint for the key event definition. He pointed out that this key event could be break down into two states, ``\textit{when the ball is on top of the hand}'' and ``\textit{when the ball is thrown into the air.}'' To indicate that the ball is thrown, ``\textit{we could use a change in the relative direction of the ball and the player's head}.''

\textbf{Key frame Marking}. Using the frame marking functionality within the \textit{Dataset View}, Alex marked two frames by clicking the button, and two markers were displayed on the timeline, as shown in \autoref{fig:datasetview}C1. These two frames represent the ``ball held by hand'' and ``ball thrown into the air,'' respectively, which correspond to the two states for later reference. At this point, he noticed that in the table tennis broadcast videos, the visual features differ significantly when the serve player is oriented to the camera versus away from it. Alex proposed that two key events be defined, and he decided to ``\textit{first define the one for the serve action of players oriented to the camera.}''

\begin{figure*}[ht]
  \centering
  \includegraphics[width=\linewidth]{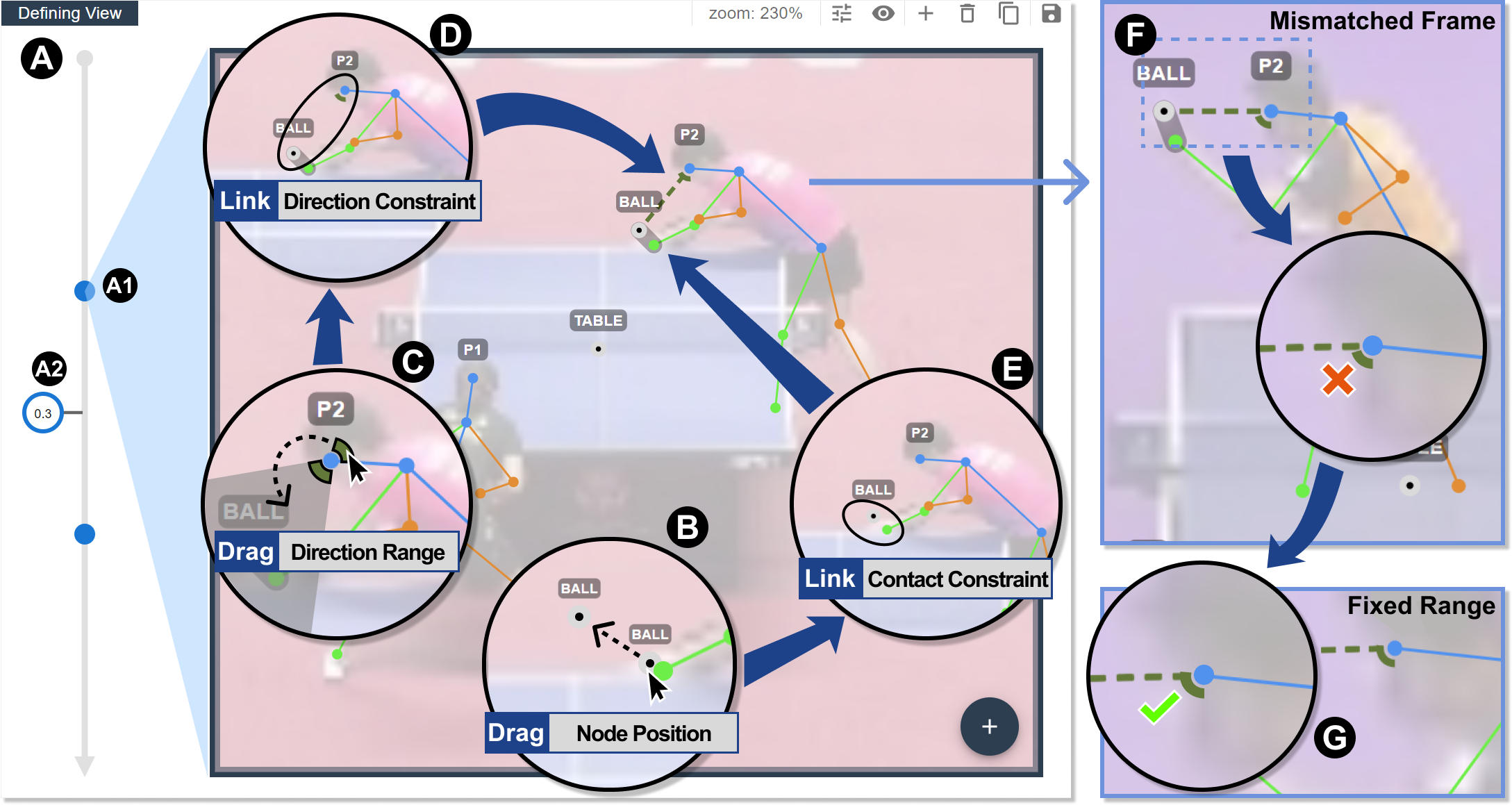}
  \caption{The \textit{Defining View} contains: a timeline (A) for setting the number of states and time intervals, and a canvas featuring drag-and-link interactions. Users can drag to adjust node positions (B) and direction ranges (C), link nodes to define constraints such as direction (D), distance, and contact (E). This design also facilitates the refinement of key event definitions (F).}
  \Description{This figure shows the drag-and-link interaction design of \framework{}, users can drag to adjust node position and direction range, link to set constraint such as direction, distance and contact. Additionally, users can interactively refine key event definitions.}
  \label{fig:defview}
\end{figure*}

\subsubsection{Defining of Key Event}

With the concept in mind, Alex proceeded with the defining. For convenience, we will refer to this key event as $K_1$ below.

\textbf{Creation of Key Events and States}. Alex initially created a new key event and initiated the editing process within the \textit{Defining View}. Subsequently, within the \textit{Defining View}, a timeline component (\autoref{fig:defview}A) was utilized for the purpose of managing the state of key events. Each node on the timeline represents a discrete state (\autoref{fig:defview}A1). Alex created two blank states, following \autoref{equ:eventtostate}, and started editing the initial one.

\textbf{Visual Element Manipulation}. The visual elements involved in the state should be set according to \autoref{equ:elementinevent} and \autoref{equ:elementattr}. The \textit{Defining View} supports two methods for visual element adding. The first method is by category, where users can select and add one visual element at a time to the canvas. The second method is through a hot start, allowing users to select a frame from any video and import all the extracted visual elements to the canvas based on their positions in the frame. Additionally, the selected frame can be set as the background of the canvas for reference. Each visual element, including objects and body parts, is represented as a node on the canvas, with the human skeleton also displayed (P3).

Alex remarked, ``\textit{Adding elements needed one by one is tedious. I've already marked some frames, so it'll be quicker to use those for a hot start.}'' He then used the second method to add visual elements, locating the previously marked frame where ``the player holds the ball and prepares to throw it up,'' and imported both the visual elements and the frame into the canvas. Alex then removed unnecessary elements, such as spectators. Since the nodes representing the hand and the ball were too close together, making them overlapping and difficult to select and link, Alex dragged the two nodes to adjust their position to separate them (P2, \autoref{fig:defview}B). It is noteworthy that the absolute position of visual elements is not a constraint and will not be considered in the final rules that generate training labels.

\textbf{Constraint Setting}. Alex then began setting the constraints between visual elements (P5, \autoref{equ:relationinevent}, \autoref{equ:relationattr}). For state 1 of $K_1$, Alex explained, ``\textit{To capture the state where the ball is still in the hand and hasn't been thrown, there are two key relations: the contact between the ball and the hand and the direction of the ball relative to the head.}'' He set the contact relation by clicking to link the ball and the hand (E5), with the relation visualized on the canvas (\autoref{fig:defview}E). For the direction relation, Alex created a valid direction range on the head node, visualized as a thick arc with the node at its center, and the arc's central angle representing the specified range. By dragging the arc, he adjusted its orientation (\autoref{fig:defview}C) and linked it with the ball node (\autoref{fig:defview}D), thereby establishing a direction constraint within a 70-degree interval toward the lower left (E5). Next, Alex established the direction relation between each player and the table. ``\textit{This relation is important,}'' he noted, ``\textit{because in this key event, the player serving the ball should be positioned above the table, while the other player should be below it.}''

Alex then began defining state 2 of $K_1$. ``\textit{For state 2, I need to set the relation between the ball and the serve player's head,}'' he explained. ``\textit{At this point, the ball is thrown up, positioned above and to the left of the center of the head.}'' He configured this in the \textit{Defining View}.

\textbf{State Interval Setting}. Referring to the previously marked frames, Alex set the time interval threshold (\autoref{equ:interval}) between the two states to 0.3 seconds on the timeline component (\autoref{fig:defview}A2).

\subsubsection{Iterative Key Event Definition Refinement}

At this point, Alex felt that his definition of $K_1$ had ``\textit{reached a temporary conclusion.}'' He decided to ``\textit{check the quality of the labels first.}'' After clicking the button, the system generated labels and displayed them in the \textit{Dataset View} (P6).

\textbf{Label Review}. In the cell matrix component, each video cell is color-coded based on the number of labels (\autoref{fig:datasetview}A1). When viewing a video, the auxiliary timeline above the progress bar displays dots indicating the distribution of labels (\autoref{fig:datasetview}C2). Alex began by selecting a few cells to review the labeled frames in the corresponding videos. Concurrently, he utilized the \textit{Frame View} to observe the retrieved frames that were based on the rules of the current state in \textit{Defining View}. Alex noticed that ``\textit{most of the labels are correct, but there are some mislabeling and missing issues.}''

\textbf{Iterative Modification}. ``\textit{I want to see why this frame wasn't labeled,}'' Alex remarked. He replaced the background in the canvas with the frame that wasn't retrieved and compared it with the previously defined relations on the canvas (\autoref{fig:defview}F). ``\textit{Ah, the range of the direction angle I set was a bit too narrow.}'' He then adjusted the angle range to encompass the direction angles in several frames that should have met the conditions (P6, \autoref{fig:defview}G). Alex made several similar adjustments until he was ``\textit{basically satisfied}'' with the results. ``\textit{I should address the mislabeling issue now,}'' he said as he began reviewing frames that were incorrectly labeled. He discovered that some frames were mislabeled due to interference from other people in the video. In state 2, since only the direction of the ball relative to the head and the direction of the person relative to the table were constrained, those frames ``\textit{meet the rules when matching the person nearby.}'' To filter out this issue, he added a pair of distance constraints. 

Alex made several more modifications to improve the quality of the label. ``\textit{That's good enough,}'' he said, deciding to stop making further adjustments. ``\textit{Even though the labels aren't completely accurate and some instances were still missed, from a data programming perspective, this is within a reasonable range.}'' Similarly, Alex defined the key event, denoted as $K_2$, which pertains to the serve action of the player oriented away from the camera. Ultimately, Alex completed the label generation in 26.6 minutes.

\subsubsection{Final Training of the TAL Model}

Alex clicks the button in the \textit{Training View} (\autoref{fig:screenshot}E) to initiate TAL model training using the labels generated by the most recent version of the two defined key events.

\textbf{Training Status}. Alex observed the training process through the \textit{Training View}, which illustrates the alterations in loss and mAP (calculated based on several labeled ground truths) as the number of training epochs increases. After the training converged, Alex acquired a TAL model for localizing serve actions and expressed satisfaction with the model performance.

\begin{table*}[t]
  \centering
  \caption{Model performance comparison with fully supervised method and single-frame supervised method.}
  \begin{tabular}{ccccccccc}
      \toprule
      \multirow{2}{*}{} & \multicolumn{7}{c}{mAP@tIoU} & avg-mAP\\ \cmidrule{2-9}
       & 0.1 & 0.2 & 0.3 & 0.4 & 0.5 & 0.6 & 0.7 & 0.1:0.7\\
      \midrule
      SF-Net w/ full label & 1.000 & 0.992 & 0.953 & 0.897 & 0.834 & 0.663 & 0.494 & 0.833\\
      SF-Net w/ single-frame label & 0.982 & 0.909 & 0.836 & 0.767 & 0.613 & 0.327 & 0.116 & 0.650\\
      \framework{} & 0.909 & 0.909 & 0.892 & 0.888 & 0.854 & 0.728 & 0.596 & 0.825\\
      \bottomrule
  \end{tabular}
  \label{tabletennis_eval}
\end{table*}

\subsection{Evaluation of the Framework}

\textbf{Comparative Study.} We conducted a comparative study to evaluate the effectiveness of \framework{} by comparing the performance of a model built using \framework{} with models trained using traditional annotation-training workflows. We manually annotated Alex's videos for training (took a total of 15.7 hours to annotate) and an additional 130 videos for testing. First, a model was trained using SF-Net with full supervision labels, followed by another model trained with SF-Net using single-frame supervision labels. For single-frame labels, we selected the central frame of each action instance. Then these two models were compared with the model Alex built using \framework{}. As shown in \autoref{tabletennis_eval}, Alex's model significantly outperformed the single-frame supervised SF-Net in terms of average mAP and approached the performance of the model trained with full supervision. This outcome is impressive given that the labeling time was reduced by over 30 times. At IoU thresholds ranging from 0.3 to 0.7, Alex's model achieved higher mAP scores than the single-frame supervision model, showcasing its robustness in modeling action duration. These results demonstrate the effectiveness of \framework{} in constructing TAL models from unlabeled video dataset.

\begin{figure*}[t]
    \centering
    \includegraphics[width=\linewidth]{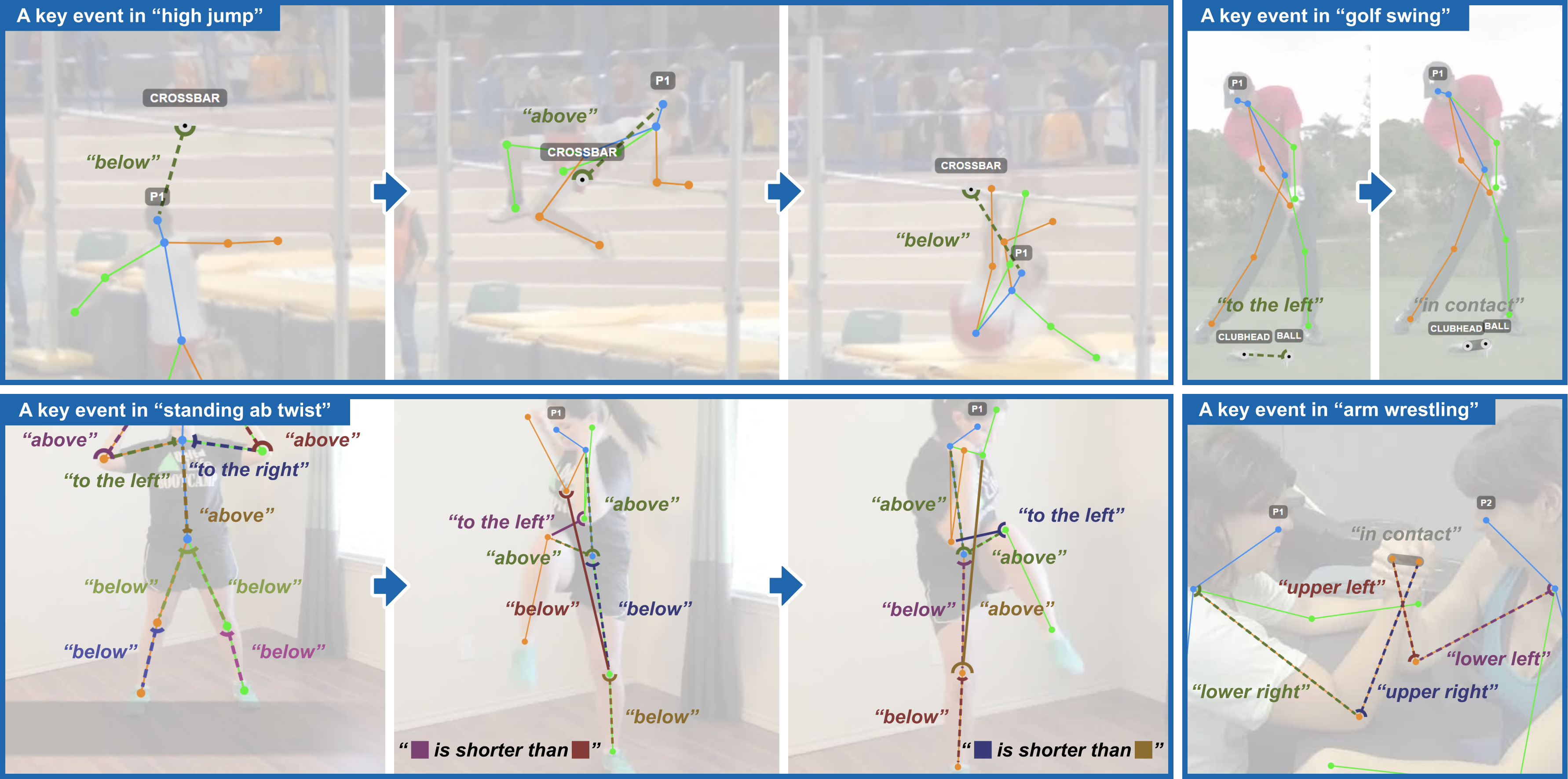}
    \caption{More usage examples. For ``high jump,'' users can consider the direction between the head and the crossbar; for ``golf swing,'' users can examine the direction and contact relation between the clubhead and the ball; and for ``arm wrestling'' and ``standing ab twist,'' users can focus on the direction relation and relative distance between the joints.}
    \Description{This figure shows examples of defining key events with our system. The four actions are high jump, golf swing, arm wrestling, and standing ab twist.}
    \label{fig:gallery}
\end{figure*}

Additionally, \framework{} can be applied to various types of actions. \autoref{fig:gallery} presents screenshots that demonstrate the use of \framework{} to define key events for different actions, including single-human actions, human-human interactions, and human-object interactions.

\section{User Study}

The effectiveness of the framework in building TAL models from unlabeled video dataset was demonstrated in \autoref{sec:ui}. To further evaluate the design of the drag-and-link interactions, we conducted a comparative user study\footnote{The study has been approved by State Key Lab of CAD\&CG, Zhejiang University.}. This study compared the drag-and-link interface of the prototype system with a baseline system that uses a form-based interface, which can be considered a version of \framework{} without the drag-and-link feature. It aimed to answer the following two research questions:
\begin{itemize}[leftmargin=*]
\item Can drag-and-link interactions reduce the time consumed in defining key events (improve the efficiency of TAL data programming)?
\item Can drag-and-link interactions reduce the number of iterations to refine key events (help define key events accurately)?
\end{itemize}

\subsection{Participants}

A total of 12 action annotators (A1–A12, Age: 22–28) were recruited for the study, comprising both male and female participants. The participants are data analysts for various sports, with extensive expertise in annotating action data for purposes including quantitative analysis, visualization, and model training. Specifically, for model training purposes, eight participants had previously engaged in annotation for more than three projects, while four had engaged in at least one. All participants understood the TAL task setting and the deep learning-based TAL model training workflow, enabling them to provide valuable insights into the framework and system. They had no involvement in the preceding formative studies. For the subsequent tasks, the participants were randomly divided into two groups (G1 and G2), with six participants in each group. The study was conducted on a PC with a 32-inch monitor in the laboratory, and each participant received a compensation of \$15.

\begin{figure*}[t]
    \centering
    \includegraphics[width=\linewidth]{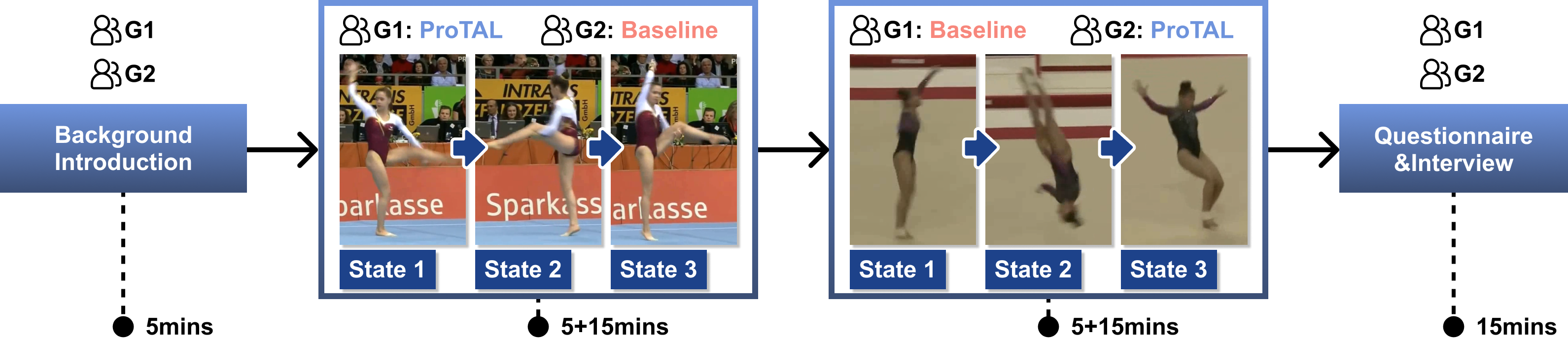}
    \caption{\textbf{User study procedure.} The user study was conducted in three phases. The initial phase comprised a background introduction. The subsequent phase involved the completion of two tasks, the first of which defined key events of ``turns'' and the second of which defined key events of ``tumbling'' in floor exercise. Participants were required to complete the tasks using different systems according to their group. Finally, a post-task questionnaire and an interview were conducted.}
    \Description{This figure shows the procedure of the user study. The first phase was an overview of the the background. The next phase comprised two tasks that needed to be finished. The first task is to define key events for ``turns,'' while the second task for ``tumbling''. Participants were asked to finish the tasks using different systems based on their group. The last phase involved a post-task questionnaire and an interview.}
    \label{fig:procedure}
\end{figure*}

\begin{figure}[ht]
  \centering
  \includegraphics[width=\linewidth]{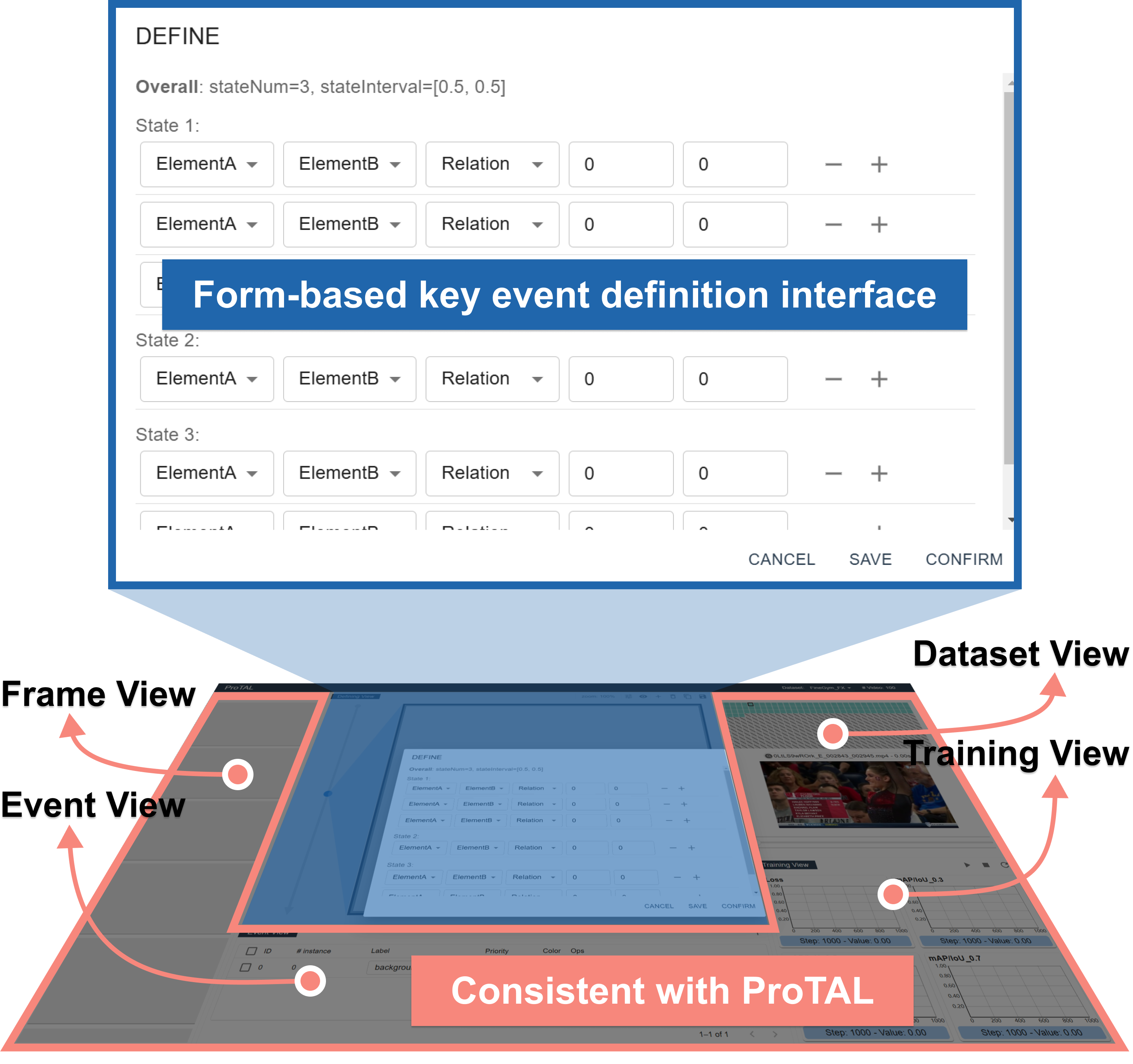}
  \caption{The baseline system employs a form-based interface for key event definition.}
  \Description{This figure shows the interface of the baseline system, which employs a form-based interface for key event definition.}
  \label{fig:baseline}
\end{figure}

\subsection{Procedures}

To assess how effectively \framework{} helps users translate abstract key event concepts into accurate, rule-based definitions, we compared it with a baseline system.
Unlike \framework{}'s drag-and-link interface, the baseline employs a form-based interface for key event definition, as shown in \autoref{fig:baseline}. In this interface, each row represents a constraint, structured as a 5-tuple: (Element A, Element B, Relation Type, Parameter 1, Parameter 2). Users set constraints by selecting visual elements and relation types from drop-down menus and inputting relevant parameters. For instance, to define a direction relation such as ``the angle of Element A relative to Element B is between 60 and 120 degrees'', the users select ``Element B,'' ``Element A,'' and ``direction,'' then specifies the angle range by entering the numerical values ``60'' and ``120.''

We designed four tasks, each involving the defining of key events in a floor exercise action (either ``turns'' or ``tumbling'') using one of the two systems (\framework{} or baseline). So, the four tasks were: (1) \textbf{\framework{}-turns}, (2) \textbf{baseline-turns}, (3) \textbf{\framework{}-tumbling}, and (4) \textbf{baseline-tumbling}. These two single-human actions were selected for two reasons: first, to reduce the effort required to understand the key events so participants can focus on the interaction and second, to ensure a comparable level of complexity in defining key events for both actions. Moreover, since the interactions designed for human- and object-related visual elements are identical, the study results are expected to remain consistent across different action types. Group G1 was asked to complete tasks 1 and 4 in sequence, while group G2 was assigned tasks 2 and 3 in sequence. This ensured that each participant experienced both systems and different actions for each system. This was necessary because defining the same key event with another system would introduce bias. Additionally, the system order was alternated between groups to maintain fairness in comparison.
The study for each participant was comprised of three phases (\autoref{fig:procedure}):

\textbf{Phase 1. Background Introduction} (5mins). The first phase involved introducing key concepts to explain how data programming works for TAL, ensuring that participants developed a comprehensive understanding of the key event and the distinction between traditional data annotation and the data programming paradigm, thus preparing them for the tasks ahead.

\textbf{Phase 2. Two Tasks} (40mins). Each participant was required to complete two tasks in sequence. Before each task, we introduced the system and the action involved in the task. Participants were shown an video collected from FineGym~\cite{shao2020finegym} containing at least two instances of the action, with the action locations marked in the timeline. To minimize the impact of varying levels of participants' familiarity with the actions, we simplified the data programming task. First, participants familiarized themselves with the action in the video, after which we provided a general description of the key event directly. For ``turns'' action, the two instances in the video involved the athlete lifting her left leg to a near-horizontal position while rotating her body. For the ``tumbling'' task, the key event was the change in the athlete's torso direction. Participants were then asked to define the key events and generate labels on the given video using the assigned system.
Each task was considered complete when the generated labels meets a specified quality (accuracy $\geq$ 0.8 and recall $\geq$ 0.2). To ensure balanced effort, the completion time ($\leq$ 15 minutes) and the number of iterations ($\leq$ 5) were capped to prevent participants from overthinking or defining key events too casually.

\textbf{Phase 3. Post-task Questionnaire and Interview} (15mins). After completing the two tasks, participants were asked to fill out a questionnaire to rate their experiences with the two systems. Following this, an interview was conducted to gather detailed user feedback on the two systems.

\subsection{Research Data Collection and Analysis}

To address the formulated research questions, a diverse set of data was collected for analysis, encompassing both quantitative and qualitative, as well as subjective and objective measures. Subjective data included questionnaire responses and interviews from the 12 participants.

\textbf{Questionnaire.} The questionnaire comprised two main sections. The first section focused on a comparative evaluation of the two systems. To explore whether the drag-and-link interaction enhances usability, six key aspects were derived from the ten questions in the System Usability Scale (SUS). Participants rated these aspects on a comparative 7-point scale ranging from ``\textit{System A (baseline) much better}'' to ``no difference'' to ``\textit{System B (our system) much better}.'' The aspects included: \textit{overall performance}, reflecting the user's overall experience with the system; \textit{easy to use}, indicating the ease of use of the system; \textit{intuitive}, assessing functional consistency and learning intuitiveness; \textit{cognitive effort}, measuring the cognitive load imposed on the users by the system; \textit{physical effort}, evaluating the physical burden during operation; \textit{practicality}, determining whether the system meets the user's practical needs.
The second section required participants to rate specific interaction designs in our system, focusing on node (visual element) manipulation and constraint setting. These ratings were collected using a 7-point Likert scale. All responses were gathered for subsequent statistical analysis.

\textbf{Interview.} The interview focused on three topics: 1) the strengths and weaknesses of the two systems; 2) the underlying reasons behind participants' behaviors that differed from others during the tasks; and 3) suggestions for improving the system and framework. All interviews were recorded and transcribed for analysis. Feedback was categorized according to the research questions and interview topics, then reviewed and discussed by three coauthors. Key results were subsequently summarized.

For objective data, we recorded the entire process of the 12 participants performing the tasks and extracted relevant data metrics for analysis. Given that the label quality was required to meet predefined standards, the metrics analyzed focused on two aspects: task completion time and the number of iterations required to complete each task. These two metrics correspond to the two research questions.

\textbf{Completion Time.} To analyze task completion times, we used a paired t-test to compare the differences in completion times for the tasks completed on the two systems by the same group of participants. First, we ran a Shapiro-Wilk test at a significance level of 0.05 to check the normality of the paired differences. Given that the differences followed a normal distribution, we calculated the mean and standard deviation for both sets, along with the t-value and p-value for the comparison.

\textbf{Number of Iterations.} The number of iterations refers to the total number of times users generated labels based on the defined key events and review labels for refinement until the required label quality was achieved. The analysis for the number of iterations followed the same procedure as that used for task completion time.

\begin{figure*}[t]
    \centering
    \includegraphics[width=\linewidth]{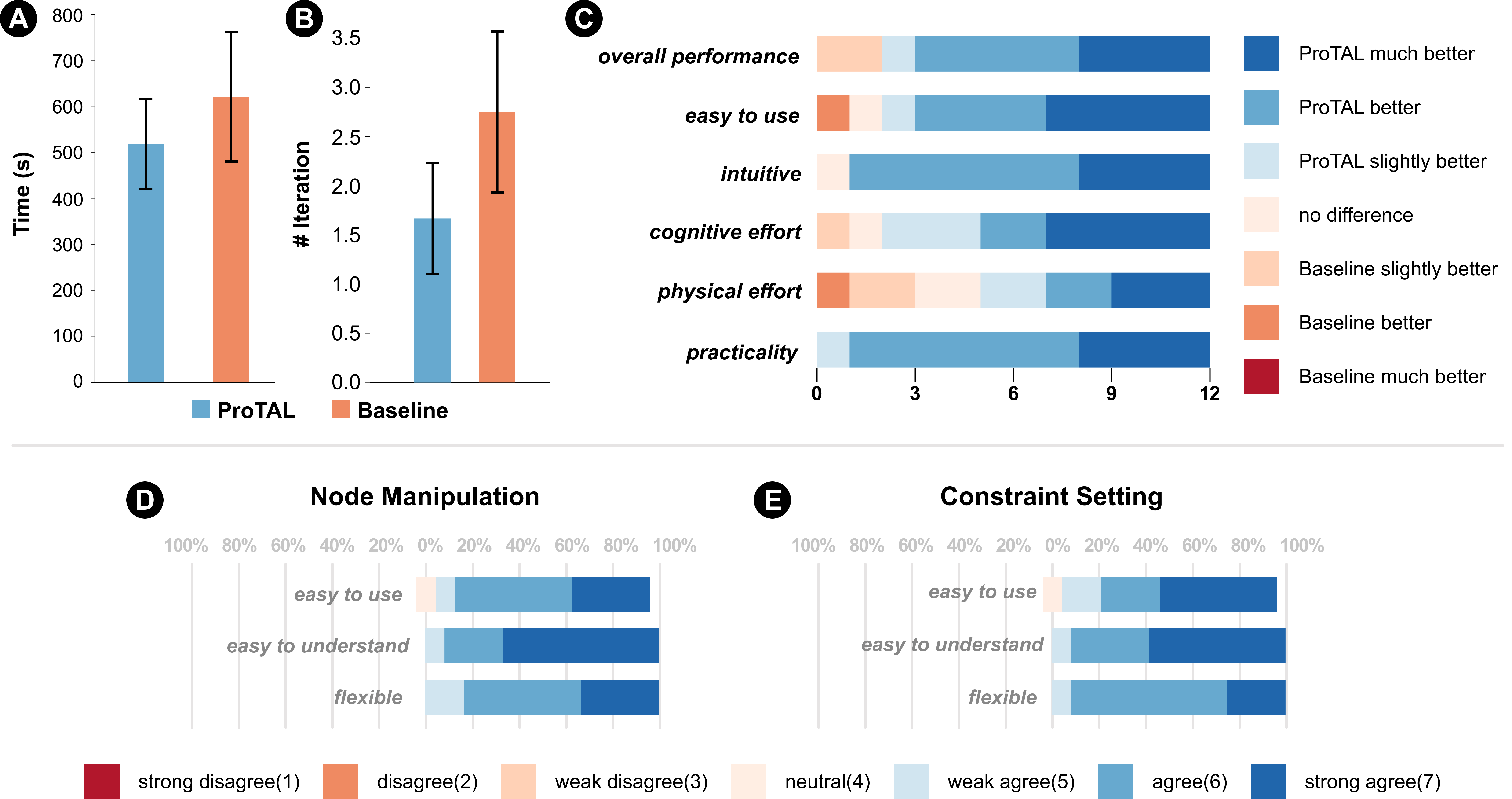}
    \caption{Quantitative results. (A) and (B) show the mean values of task completion time and the number of iterations, respectively, with error bars indicating the 95\% confidence interval. (C) presents users' overall comparative ratings of our system versus the baseline. (D) and (E) display ratings for the interaction design of node manipulation and constraint setting in our system, respectively.}
    \Description{Five charts show the quantitative results of our user study. The first and second are bar charts depicts the mean values and the error bars represent the 95\% confidence interval, shows the completion time and number of iterations in tasks with our system and baseline. The third one is shows the comparative ratings of our system and baseline. The last two charts display user ratings for the interactive design of node manipulation and constraint setting respectively.}
    \label{fig:quantres}
\end{figure*}

\subsection{Results and Feedback}

\subsubsection{Drag-and-link interaction enhances usability compared to the baseline system}

Participants' ratings of the overall comparison between two systems are presented in \autoref{fig:quantres}C. For overall experience, 83.3\% (10/12) of participants preferred our system over the baseline, while only 16.7\% (2/12) rated the baseline as ``slightly better''. Regarding interaction and visual encoding, 10 participants found our system ``easier to use'' and ``more intuitive,'' whereas 2 participants, A4 and A10, who were highly familiar with gymnastics and the given actions, confidently defined the key events and found the baseline smooth to use. A4 noted that since he already knew exactly how to define the key event and what angle to set, he did not need the drag-and-link feature.

In terms of cognitive effort, 83.3\% (10/12) of participants felt that our system required less cognitive effort. A1 and A12 noted that translating the desired direction into a numerical representation with the baseline was cognitively demanding. A11 observed that using the baseline often resulted in setting angles ``\textit{based on intuition and not sure.}'' A7 emphasized that handling complex actions would be challenging with the baseline. Conversely, A2 and A4 mentioned that with a deep understanding of the action, the baseline could also be used effectively.
Regarding physical effort, there was no clear preference between the two systems. A2 noted that the baseline interaction was also straightforward, the primary limitation of it is its lack of ability to facilitate interactive exploration.
For practicality, all participants expressed a preference for using our system in their practical workflows compared to the baseline. These findings demonstrate that our system is more usable than the baseline, due to the introduction of the drag-and-link interaction.

\subsubsection{The interaction design for node manipulation and constraint setting are intuitive}

Most participants provided positive feedback on the manipulation of nodes representing human- and object-related visual elements, as shown in \autoref{fig:quantres}D. However, there was a neutral rating regarding ease of use, with A2 suggesting, ``\textit{When dragging the entire human skeleton as a whole, it is easier to use and understand using a key combination to distinguish it from dragging individual nodes.}'' Regarding constraint setting, most participants also gave favorable ratings for the interaction design. A1 highlighted the usefulness of the mode switch feature (\autoref{fig:screenshot}C1), noting that when the angle range is narrow, such as 15 degrees, the arc representing the range is small, making it somewhat challenging to drag. However, by switching the display mode, this issue was effectively resolved. Additionally, all participants found that setting a frame as the canvas background was highly useful, as it shows placement of visual elements and provided intuitive cues for setting constraints.

\subsubsection{Drag-and-link interaction enhances efficiency in TAL data programming}

Regarding the first research question, a paired t-test comparing task completion times between the two systems revealed that participants completed tasks faster using our system compared to baseline ($t = 3.04$, $p < 0.05$). The average completion time with our system was $518.8$s ($SD=172.8$), whereas the baseline required $622.1$s ($SD=249.4$), shown as \autoref{fig:quantres}A. These results demonstrate that the drag-and-link interaction improves efficiency in TAL data programming. Additionally, both times were well within the allotted 15 minutes ($900$s), indicating that participants were able to understand and adapt to the data programming workflow and the constraint-setting logic with ease. Furthermore, All participants highlighted our system's ability to directly compare the defined constraints with a selected frame, which allows them to modify the constraints more efficiently to filter or retrieve some frames. A1 and A3 emphasized that the constraint copy feature also speeds up the process, noting that constraints are often similar between states.

\subsubsection{Drag-and-link interaction helps define key events accurately}

The paired t-test revealed a significant reduction in the number of iterations required when using our system compared to baseline ($t = 2.60$, $p < 0.05$). In this study, the maximum number of iterations was capped at 5, with values ranging from 1 to 5. The average number of iterations for tasks completed with our system was $1.7$ ($SD = 0.9$), compared to $2.8$ ($SD = 1.3$) with the baseline, as shown in \autoref{fig:quantres}B. These results indicate that participants required fewer iterations to complete the key event definitions using our system than with the baseline.
With our system, 50\% (6/12) of the participants completed the task in a single iteration, while only 25\% (3/12) achieved this using the baseline. This suggests that the drag-and-link interaction enables users to define more accurate rules in the initial iteration. Furthermore, excluding cases with only one iteration, the average number of iterations with our system was $2.3$, compared to $3.3$ for the baseline, highlighting that the drag-and-link interaction facilitates more precise rule modifications. These findings show that the drag-and-link interface provides a more accurate approach to define key events, answering the second research question.

\subsubsection{Individuals displayed a diversity of patterns of behavior and cognitive processes}

During the tasks, notable diversity was observed in the way participants defined key events. For example, in terms of constraint setting, A1 and A2 initially set a wide angle range for direction constraints and then narrowed it in subsequent iterations. In contrast, A3 and A6 took the opposite approach. A1 indicated a preference for initially relaxing the constraints and then tightening them to eliminate incorrect frames, while A2 emphasized the importance of ensuring a high recall rate at the beginning. In contrast, A3 and A6 prioritized accuracy and then sought to improve recall. From the perspective of visual element selection, A6 and A12 found and attempted to define different versions of the key event definitions for the tumbling action. They focused on the direction of the person's feet and head, defining more complex but effective key events. This phenomenon is consistent with the nature of key events, where different users may have different interpretations of the same action. For any action, there may be several reasonable key events to define.

\subsubsection{The system exhibits significant potential for improvement}

During the interviews, participants agreed that \framework{} offers a promising solution to the high cost of action annotation and provided several suggestions for improvement to address its perceived weaknesses and improve usability.
Three participants expressed concern about the numerical accuracy of the constraints, especially since they had previously annotated precise data. They recommended combining drag-and-link interaction with direct numerical control in the baseline system to enhance numerical accuracy.
A8 suggested implementing an adsorption effect to adjust the direction range to improve interaction efficiency and precision.
Currently, our system highlights the generated labels in the \textit{Dataset View}, allowing users to review and adjust constraints for mislabeled frames. A1 suggested further refinements, such as highlighting specific regions within the mislabeled frames that do not meet the defined constraints. This feature would eliminate the need for users to manually identify which constraints caused errors. A7 recommended that the system automatically update constraints based on mislabeled frames.
In addition, A1, A7, and A8 suggested that the integration of language models could significantly improve the efficiency of labeling by recommending potential constraint candidates when defining key events.

\section{Discussion}

In this section, we reflect on our interactive video programming framework and the prototype system, summarizing the implications we learned. We also discuss the feasibility of \framework{}, explore possible future research directions, and outline the limitations of current research based on user feedback and observations.

\subsection{Implications for Designing Video Programming Framework}

The effectiveness of \framework{} and the usability of the prototype system are demonstrated, highlighting the potential to inspire the design of data programming frameworks for other video tasks.
\begin{itemize}[leftmargin=*]
    \item \textbf{Identify the appropriate constraint space for new video programming tasks}. In this paper, our goal is to develop a video programming framework for TAL. We began by decomposing actions into finer-grained key events, defining them through changes in the relations between visual elements, which serve as labeling functions in data programming. To better understand the constraints involved in defining key events, we conducted a workshop study that led to the derivation of the constraint space. This space guided the implementation of the prototype system, which was successfully applied to practical scenarios. However, this constraint space may not encompass all video tasks. When designing video programming frameworks for other tasks, it is crucial to carefully derive the constraint space for them. For instance, when developing a system for higher-level event recognition, such as tactical analysis in team sports~\cite{lza}, the constraint spaces should be extended to encompass lineup information, player roles, etc.

    \item \textbf{Decomposing and simplifying data programming objects for highly complex tasks}. Data programming is being applied to increasingly complex tasks and data, moving from text to images and from video classification to action localization. However, the complexity that rules can handle is not keeping pace with the growing complexity of tasks and data. In addition, the rules must remain simple enough, as overly complex rules would make direct annotation more efficient than data programming. Therefore, when extending video programming to more complex tasks, it is essential to decompose complicated programming objects, such as decomposing actions into key events with a simpler structure and programming key events. Such decomposed objects can be defined by rules of manageable complexity, facilitating data programming. Furthermore, advanced models are needed to use these weak labels for effective model training.
\end{itemize}

\subsection{Potential of \framework{}}

We reflect on the design and potential of the framework and system, focusing on adaptability and scalability.
\begin{itemize}[leftmargin=*]
    \item \textbf{Adaptability to broader applications.} Although \framework{} is specifically designed for TAL, its drag-and-link interaction design, along with its visual encoding of visual elements, human poses, and constraints between them, can be extended to other tasks involving actions or interactive events, such as action quality scoring and spatial action segmentation.
    \item \textbf{Efficiency in handling larger datasets.} \framework{} effectively scales with dataset size without increasing annotator workload. The annotator's time cost remains consistent as the dataset size grows, since they only need to define key events. \framework{} then automatically applies these definitions to match all frames, eliminating the need for additional manual intervention. This efficiency makes \framework{} a viable tool for large-scale video datasets.
    \item \textbf{Extension to higher-dimensional scenarios.} Although currently focused on video data, \framework{} can be expanded to handle 3D~\cite{CHEN202111} or even 4D scenarios, such as those in virtual reality~\cite{LOPES202413, ZHANG202322} and motion capture systems. By incorporating 3D detection or tracking modules, the system's canvas can be extended to define key events in the 3D space. This extension opens opportunities for annotating complex interactions and actions within immersive environments.
\end{itemize}

\subsection{Limitations \& Future Work}

\subsubsection{Current limitations of \framework{}}
While \framework{} has proven effective for temporal annotation across various types of actions, it may face challenges in complex in-the-wild scenarios. Firstly, dense and overlapping objects in videos can complicate the recognition and extraction of visual elements, thereby disrupting the data programming workflow. Changes in viewpoint present another challenge. In cases where the video dataset features distinct viewpoints, such as the two viewpoints in the table tennis match videos discussed in \autoref{sec:ui}, users can define separate key events for each viewpoint. However, dynamic or excessively varied viewpoints may require viewpoint alignment or defining key events within a 3D environment to ensure consistency. Additionally, videos shot from a first-person perspective introduce unique complexities, such as handling the hands, body, or other visible parts of the shooter, which may require tailored approaches. Further exploration will be conducted to address these limitations.

\subsubsection{Future work}
Moreover, there are several opportunities for future work:
\begin{itemize}[leftmargin=*]
    \item \textbf{Expanding the space of constraints for greater flexibility}. Currently, \framework{} provides a set of constraints based on the relations between visual elements. However, in order to distinguish actions in a more fine-grained way, such as distinguishing between tumbling actions on the ground and in the air, \framework{} needs to support a larger constraint space. This extension would allow users to define more nuanced actions and handle complicated action variations, further improving the accuracy and flexibility of action annotation.

    \item \textbf{Domain knowledge-driven constraint recommendation}. Our user study has shown that users may have different cognitive understandings of actions, and users who have a deeper understanding of specific actions can define the key event more efficiently. Therefore, \framework{} can benefit from integrating domain knowledge to automatically recommend appropriate constraints based on the specific action. By integrating expertise in different actions, \framework{} can guide users to select constraints that are more appropriate for their tasks, thus reducing cognitive load and improving the accuracy of the annotation process.

    \item \textbf{Integration with large multimodal models}. Incorporating large multimodal models into \framework{} could enable more advanced AI-powered features. Using video, image, and text data, multimodal models could automatically suggest key events and constraints based on the context of the action, simplifying the process of defining key events. Furthermore, large multimodal models offer the potential to integrate \framework{}'s visual element extraction and rule-based frame matching steps. For example, users could enter rules in natural language along with a frame, and the models could determine whether the frame satisfies those rules, further increasing flexibility.
\end{itemize}

\section{Conclusion}

We present \framework{}, a novel video programming framework designed for TAL. The framework addresses the significant challenge of decomposing actions into meaningful substructures by decomposing actions into key events that are easier to define and recognize. \framework{} then presents a drag-and-link interaction design that allows users to define key events through intuitive interactions. These key event definitions, which constrain relations between visual elements, serve as data programming rules that generate frame-wise action labels for large-scale unlabeled videos. With these labels, a semi-supervised method is used to effectively train TAL models.

Based on the proposed framework, a system was implemented. The effectiveness and usability of the implemented system in TAL annotation and training was demonstrated through a practical usage scenario and a user study. Feedback from participants highlighted the design of the drag-and-link interaction. These results also provide valuable guidance for the development of future video programming frameworks.

\begin{acks}
This work was supported by NSFC (U22A2032) and Key Scientific Research Project of the Department of Education of Guangdong Province (2024ZDZX3012). The author also gratefully acknowledges the support of Zhejiang University Education Foundation Qizhen Scholar Foundation.
\end{acks}

\bibliographystyle{ACM-Reference-Format}
\bibliography{main}










\end{document}